\documentclass[12pt]{iopart}

\usepackage{graphicx}
\usepackage{multirow}
\usepackage{booktabs} % for good looking regression table
\usepackage{subcaption}
\usepackage{floatrow}
\newfloatcommand{capbtabbox}{table}[][\FBwidth]
\usepackage{soul} % for highlighting

\begin{document}

\title[The Multilayer Structure of Corporate Networks]{The Multilayer Structure of Corporate Networks}

\author{J A van Lidth de Jeude$^1$, T Aste$^2$, G Caldarelli$^{1,3,4}$}

\address{$^1$ IMT School for Advanced Studies Lucca, Piazza San Francesco 19, 55100, Lucca, IT}
\address{$^2$ Computer Science Department, University College London, Gower Street, London, WC1E 6BT, UK}
\address{$^3$ Istituto die Sistemi Comlessi (CNR) UoS Universita ``Sapienza'', P.le A. Moro 2, 00185 Rome, IT}
\address{$^4$ European Centre for Living Technology, Universita di Venezia ``Ca' Foscari'', S. Marco 2940, 30124 Venice, IT}
\ead{jeroen.vanlidth@imtlucca.it}
\vspace{10pt}
\begin{indented}
\item[]October 2018
\end{indented}

\begin{abstract}
Various company interactions can be described by networks, for instance the ownership networks and the board membership networks. To understand the ecosystem of companies, these interactions cannot be seen in isolation. For this purpose we construct a new multilayer network of interactions between companies in Germany and in the United Kingdom, combining ownership links, social ties through joint board directors, R\&D collaborations and stock correlations in one linked multiplex dataset. We describe the features of this network and show there exists a non-trivial overlap between these different types of networks, where the different types of connections complement each other and make the overall structure more complex. This highlights that corporate control, boardroom influence and other connections have different structures and together make an even smaller corporate world than previously reported. We have a first look at the relation between company performance and location in the network structure.
\end{abstract}

%
% Uncomment for keywords
%\vspace{2pc}
%\noindent{\it Keywords}: XXXXXX, YYYYYYYY, ZZZZZZZZZ
%
% Uncomment for Submitted to journal title message
% \submitto{\NJP}
%
% Uncomment if a separate title page is required
%\maketitle
% 
% For two-column output uncomment the next line and choose [10pt] rather than [12pt] in the \documentclass declaration
%\ioptwocol
%

\section{Introduction}
%Such networks have been used to show that the corporate world is a small world; few companies can exert wide control through the network structure and through the boardroom elite decisions can spread to form herd behaviour.
Networks, or graphs, are now widely used in economic and financial literature as they represent a natural way to study connections and systemic effects \cite{battiston2010structure, caldarelli2007scale}. Under corporate networks we consider all networks that describe interactions between companies. Over the last two decades, studies in the broader field of economic and financial networks studies of these systems have shown that the interconnectedness of the economic and financial system is a main driver of (in)stability \cite{May2008a,Schweitzer2009,Battiston2012DebtRank}. The topology of the networks influences the resilience to shocks \cite{Luu2018}, and the evolution of mesoscale topological structures could even indicate early-warning signals of a crisis \cite{Squartini2013a}. 

On corporate networks, \cite{battiston2003decision, davis2003small}  study spreading processes of influence on the network of interlocking board members. Through board members that work for multiple companies, decision-making spreads and this network has been shown to exhibit herd behaviour. These board interlocks can limit competition as they obstruct independent decisions by boards \cite{zajac1996director}. 

Ownership networks constructed from (partial) ownership stakes of corporate entities are another type of corporate networks. These networks have revealed a small corporate world; a large concentration of ultimate ownership in a small group of core companies in this network \cite{vitali2011network}. In \cite{rungi2017global}, ownership networks are studied to find the role of subsidiaries in control of large parent companies. 

Also innovation dynamics have been studied by networks of R\&D partnerships, showing for example the effective outsourcing of research by big corporations to start-ups \cite{tomasello2017rise}. On the financial side, corporate networks can be constructed from stock price correlations. With techniques like network backbone extraction one can identify influential companies from these networks \cite{Tumminello10421, bonanno2004networks}. These studies have shown how networks can be used to study the dynamics of competition and innovation, or identify the influence of the corporate topology one some notion of control and influence between corporations. \\

Most research until now studied these different types of corporate networks in isolation. However, corporate networks are strongly interconnected; e.g. a cascading effect between board members will influence stock market fluctuations and vice versa. We therefore argue that these systems should be studied in parallel, not in isolation. In this paper we will unravel the aggregate corporate network and characterise the topology of the different layers of this system. To this end we construct a new multilayer network of interactions between companies. We identify four main connections: \textit{ownership links, social ties through board members, research collaborations, and stock correlations}.

\begin{figure}[t!]
\centering
\includegraphics[width=\linewidth]{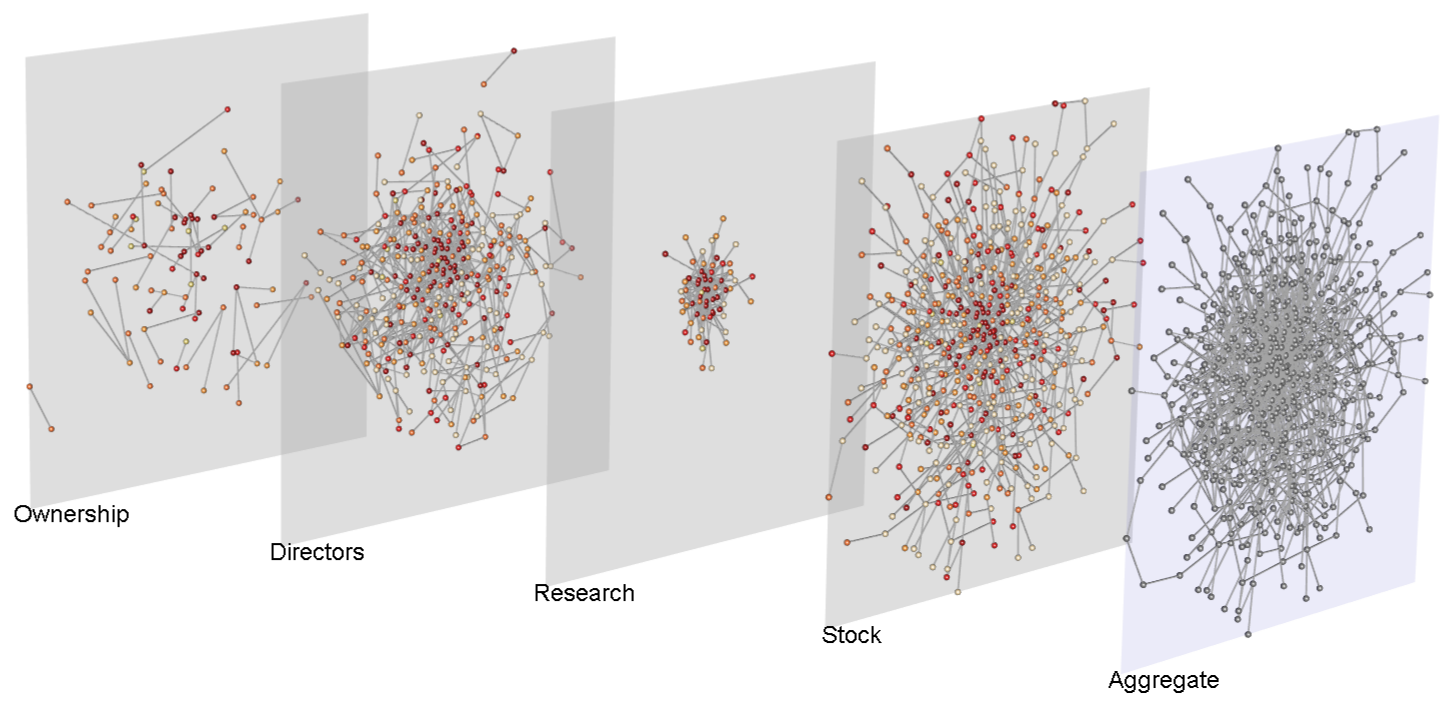}
\caption{Multilayer network of listed German companies showing: cross-stockholding ownership network, joint board directors network, research interactions, minimum spanning tree of stock correlations and the aggregate. We show a multiplex layout; the position of nodes is the same in all layers, but isolated nodes in a layer are not shown. Nodes are coloured by their MultiRank centrality in the multiplex. The visualization illustrates clearly that these interactions have different structures. All multiplex figures are created with the MuxViz software \cite{muxviz}. }
\label{fig:muxviz_german}
\end{figure}

\section{Data}
%Ownership is a clear way to exert control, simply because by law shareholders have a vote in business decisions. The board of directors make strategic decisions for the company. As board members often sit in multiple boards, this creates a channel for the spread of this decision making. A different interaction between firms comes by research collaborations. These provide a mechanism for the exchange of knowledge and innovation between companies. And finally we look at interactions of stock market relations. Networks of stock correlations are used to represent and visualise market structure \cite{mantegna1999hierarchical}.

We construct a unique multiplex dataset where we go beyond just listed companies and include all registered companies in Germany and in the United Kingdom and Ireland. The company information comes from the Amadeus database from Bureau van Dijk, this includes stock prices, ownership details, names of directors in the boards, patent data, and audit firm details. 
The data reflects the state of all registered companies in those countries as in February 2018. Daily stock prices are collected in the 2-year window before this date. 

We select companies registered in Germany with at least 10 employees (if employee data is not available this is estimated by taking the industry average based on revenue). For the United Kingdom and Ireland we study all listed companies from data that comprises 1312 companies that are listed and registered within these countries. 

Networks with links of different types are described by multilayer networks, where every layer corresponds to the interaction graph of a single type of interaction. The subset of these networks where there are no edges between nodes in different layers, is also called multiplex networks. In \cite{de2015structural}, the authors show that multilayer graph analysis in various applications gives a better representation than aggregate or single layer networks. This is driven by the fact that there are often multiple drivers of an effect, and as these are often connected but have different structures, the single or aggregate layers might under- or overestimate the network effects. 

\subsection{Networks}
\begin{itemize}
    \item \textit{Ownership.} Ownership ties are constructed from shareholder data \cite{garlaschelli2005scale}. We are interested in cross-shareholding for companies in our dataset, i.e. instances where the shareholder of a company is of itself a company in our data. We thus disregards all shareholders that are foreign or individuals. 

Ownership can be defined in various ways. Strict ownership is usually set as stake of $>50\%$. For control and influence a smaller stake (5, 10, $20\%$) is often considered sufficient \cite{imf2004directinvestment,porta1999corporate}. Peer effects of ownership can be robust under these different definitions \cite{vitali2011network}. We construct an edge between companies $i$ and $j$ when company $i$ has a stake $\geq 10\%$ in company $j$ following \cite{vitali2011network}. 

\begin{table}[t!]
\centering
\begin{minipage}{1\textwidth}
    \hfill\includegraphics[width=0.55\linewidth]{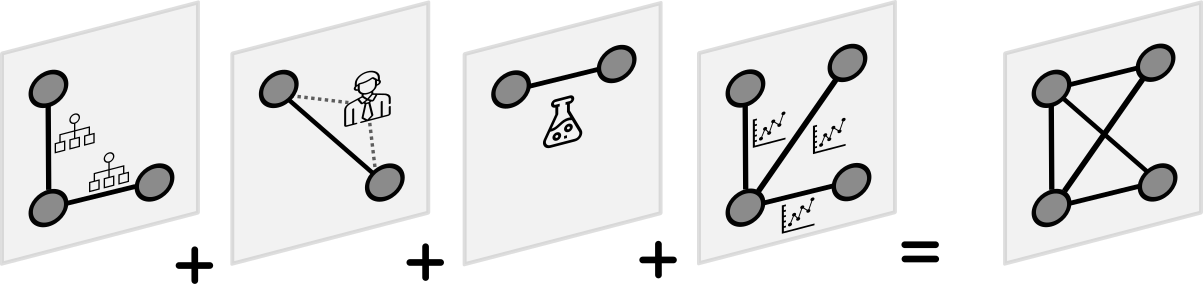}
\end{minipage}
\begin{tabular}{ll|lllll}
 & & Ownership & Board & Research & Stock & Aggregate \\ \hline
German  & Nodes & 49385 & 90690 & 3744 & & 105005\\ 
companies & Edges & 38594 & 308123 & 8420 & & 335213\\ 
& Average Degree & 1.56 & 6.80 & 4.50 & & 6.38\\ 
& Assortativity & -0.03 & 1.00 & -0.18 & & 0.98 \\
& & & & & & \\
Germany listed & Nodes & 86 & 330 & 64 & 535 & 535\\ 
companies only & Edges & 54 & 504 & 212 & 534 & 1214\\ 
& Average Degree & 1.26 & 3.05 & 6.62 & 2.00 & 4.54\\ 
& Assortativity & -0.18 & 0.48 & -0.42 & -0.03 & 0.29 \\
& & & & & & \\
UK listed & Nodes & 301 & 1043 & 25 & 1312 & 1312\\ 
companies & Edges & 293 & 2778 & 25 & 1311 & 4354\\ 
& Average Degree & 1.95 & 5.33 & 2.00 & 2.00 & 6.64\\ 
& Assortativity & -0.51 & 0.69 & -0.19 & -0.21 & -0.11
\end{tabular}
\caption{Multiplex Descriptive Statistics}
\label{table_stats}
\end{table}

\item \textit{Board of Directors.} The board membership network describes companies that are linked through directors. The board makes strategic decisions for the company. Decision making can spread through directors that sit in multiple boards \cite{battiston2003decision}. In this network an edge exists between company $i$ and $j$ when the boards of directors of company $i$ and company $j$ have a least one director in common. 
This can also be seen as the company-side projection of the bipartite network of directors and the companies they serve. We rely on unique person identifiers in the database, which prevents any problem with name disambiguation. 

\item \textit{Research.} From patent application data we obtain two kinds of links; (i) companies that are joint assignees (owners) of a patent and (ii) inventors that worked for multiple companies at the same time. The first case is a clear sign of joint research which led to the joint patent application. Following other studies on R\&D networks  \cite{hanaki2010dynamics, zacchia2017knowledge}, we also exploit inventors that appear on multiple patents with different assignee companies, i.e. inventors working with multiple companies. This can be used as a proxy of joint research \cite{cantner2006network}. Because of the long term characteristics of joint research projects, we use the patent application data between 2016-2018. 

\item \textit{Stock Correlations.} From stock price time-series we can identify significant relations between stocks. Here we follow earlier studies \cite{bonanno2004networks, mantegna1999hierarchical}, constructing a minimum spanning tree to identify a backbone network of important links. The method extracts a backbone network from pairwise correlations between the time-series. The pairwise correlations are computed on the logarithmic daily returns of a stock price. For stock $s$ with price $p$, the daily log-return is : $r_t^s = \log p_t^s - \log p_{t-1}^s$. We use a 2-year window of the stock time-series in order to have $n_s<n_t$, for a well behaved multiple pairwise correlation matrix. The 2-year window in combination with daily returns, rather than daily closing prices make the time series more comparable and diminishes effects of long term trends. In a further regression exercise we also use the full set of pairwise correlations without extracting the backbone network. 
\end{itemize}

\section{Network Topology}
In this section we first describe the network topology of the various company-to-company interactions. We then discuss the significance of the multiplex structure and show that all layers convey different structural information. 

\subsection{Layer Descriptions}
The network of German companies results in a multiplex of over a hundred thousand companies, where the layers describing different types of company interactions are all described by different networks statistics. 
For all German companies that have at least one connection to another company, we obtain an aggregate network of $105005$ companies. $89\%$ of the companies are connected through a joint board member, and slightly less than half of the companies are connected through an ownership link, see Table \ref{table_stats}. The network of research interactions is much smaller compared to the other layers, but with an average degree of 4.5, this is more densely connected than the ownership topology which has only 1.26. The largest contribution of links in the aggregate network comes from board interactions. From the assortatvity of the layers, the tendency of nodes to connect with nodes of the same size (degree), we see that while through board members companies connect very much to similar size companies, this effect is non existing in the ownership network. In the research network assortativity is negative, -0.51, driven by larger companies that collaborate with smaller companies (start-ups). 

\subsubsection{Listed Companies}
The subset of companies that are listed has similar characteristics to the full network.
For listed companies we also have stock market price interactions. When we compare the characteristics of the subset of listed companies to those of the whole system, many of the dynamics between the layers are the same. This is in principle not surprising, however, in this case the subset comprises of a particular set of the most 'important' and mainly biggest companies in terms of turnover etc. 
Nearly all companies are connected by board members, showing the potentially large influence of this more informal network. 

\subsubsection{United Kingdom}
For listed firms in the UK we observe a reasonable similarity in of the network with the German network, in terms of the assortativity of the different interactions and the relative sizes of the different layers. Main differences are a more dense board membership network, with an average degree of 5, less research interactions, as the UK economy in general is more geared towards less research intensive services than the German economy, see Table \ref{table_stats}. 

\subsection{Multilayer Structure}
In the previous section we described the topology of the isolated layers, but we are interested in these interactions in parallel. We want to understand whether we need to distinguish these different interactions, or whether the aggregate would suffice. As the multiplex literature is a very recent one, there is not one established method to evaluate this question of the relevance of the multilayer structure. Therefore we answer this with three different available methods for the analysis of the multiplex structure:
\begin{enumerate}
    \item  the node and edge overlap of the layers;
    \item  the structural reducibility;
    \item  a network regression. 
\end{enumerate} 
All these three methods on their own will show that all layers are significant and that the multiplex representation has an added value. 

\subsubsection{Node and Edge Overlap}
Node and edge overlap shows similarities between the layers and the level of connectedness. 
The node and edge overlap in two networks is measured by the fractions nodes (edges) that occur in both networks over the total number of nodes (edges). Results in Table \ref{tabel_overlap} indicate that all layers are connected but not overlapping in edges as we observe a low, but non-zero edge overlap. This holds both for the German companies and the UK companies. The fact that for all pairs of layers we low edge overlap shows that \textit{these corporate networks do not simply overlap}.

\begin{table}[t!]
\begin{scriptsize}

\begin{subtable}{.5\textwidth}\caption{\textbf{Node} overlap of German multiplex }\label{tab:1a}
{\begin{tabular}{l|cccc}
      & Ownership & Board & Research & Stock \\ \hline
    Ownership & 1 &  &  &  \\
    Board & 0.11 & 1 &  & \\
    Research & 0.02 & 0.11 & 1 & \\
    Stock & 0.16 & 0.62 & 0.12 & 1 
    \end{tabular}}
\end{subtable}%
\begin{subtable}{.5\textwidth}\caption{\textbf{Edge} overlap of German multiplex }\label{tab:1b}
{\begin{tabular}{l|cccc}
      & Ownership & Board & Research & Stock \\ \hline
    Ownership & 1 &  &  &  \\
    Board & 0.02 & 1 &  & \\
    Research & 0 & 0.06 & 1 & \\
    Stock & 0.01 & 0.03 & 0.03 & 1 
    \end{tabular}}
\end{subtable} \\
\vspace{10pt} 
\begin{subtable}{.5\textwidth}\caption{\textbf{Node} overlap of UK multiplex }\label{tab:1c}
{\begin{tabular}{l|cccc}
      & Ownership & Board & Research & Stock \\ \hline
    Ownership & 1 &  &  &  \\
    Board & 0.23 & 1 &  & \\
    Research & 0.02 & 0.02 & 1 & \\
    Stock & 0.80 & 0.23 & 0.02 & 1 
    \end{tabular}}
\end{subtable}%
\begin{subtable}{.5\textwidth}\caption{\textbf{Edge} overlap of UK multiplex }\label{tab:1d}
{\begin{tabular}{l|cccc}
      & Ownership & Board & Research & Stock \\ \hline
    Ownership & 1 &  &  &  \\
    Board & 0.002 & 1 &  & \\
    Research & 0.001 & 0.006 & 1 & \\
    Stock & 0.004 & 0.005 & 0.003 & 1 
    \end{tabular}}
\end{subtable}
\caption{\textbf{Layer overlap.} Overlap of nodes and edges between the layers, as measured by the fraction of nodes/edges which appear in both layers. Edge overlap is small but non-zero, indicating layers are complementary. }\label{tabel_overlap}
\end{scriptsize}

\end{table}

\subsubsection{Structural Reducibility}
\begin{figure}[t!]
\centering
\includegraphics[width=0.3\linewidth]{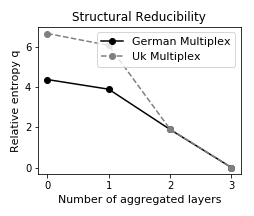}
\caption{Structural reducibility of the multiplex based on entropy of the layers. The relative entropy $q$ measures the structural information of the multilayer compared to its aggregate. The larger $q$, the more distinguishable the multiplex is from the aggregate network. When none of the layers are aggregated the entropy is maximal. I.e. none of the layers are structurally reducible based on redundant structural information measured from the entropy. For a complete description we do require all information encoded in all layers/types of interactions.  }
\label{fig:reducibility}
\end{figure}

Structural reducibility is a recently introduced measure \cite{de2015structural} for multilayer network that indicates whether pairs of layers can be aggregated based on redundant information. The information encoded in a network can be quantified by the entropy. The structural reducibility quantity calculates the relative entropy between a network of multiple layers and its aggregate. By analysing whether some layers can be aggregated, without loosing distinguishability from the aggregate, one can find the configuration of the multilayer that maximises the information in the system. The information is quantified by the entropy (Von Neumann entropy) of the network. Formally we maximise the value 
\begin{equation}
\centering
%q(G) = 1- \frac{\text{combined entropy of the multilayer}}{\text{entropy of the aggregated network}}
q(G) = 1- \case{\textrm{combined entropy of the multilayer}}{\textrm{entropy of the aggregated network}}
\end{equation}
See the Appendix for details and the formal introduction of this method.
Calculating the structural reducibility quality between all the layers, we find the the optimal structure is the multilayer with all original layers present, see Figure \ref{fig:reducibility}. This indicates that \textit{all layers convey different structural information}.

\subsubsection{Multiplex Network Regression}
A third way we can probe the significance of the different layers is by means of a regression. The multiplex network describes different types of interaction between the actors. With a regression we can test the explanatory power of these interactions for an observed interaction. For sparse networks a classic regression, where observations are formed by all possible edges (all adjacency matrix entries), one would regress with many zero entries. We use a newly proposed method more suited for network regressions that uses a graph null model to look at significant links. 
It is well known that interactions described by a network structure should be tested against a null model, to see which part of the observed network can be explained by randomness. With a null model we answer the question: out of the many combinations the network structure could possible be configured, how (im)probable is the empirically observed structure? This way we can filter observed interactions that arise from randomness from combinatorial factors. 

To this end we use a recently introduced network regression model that uses a generalised hypergeometric graph ensemble as the graph null model. This regression estimates the influence of a layer of the multiplex and, using the null model, tests the statistical significance of the layer structure on observed interactions. A detailed explanation can be found in \cite{casiraghi2017multiplex}. 
We setup a regression to see if the stock correlations can (in part) be explained by the network structures:
\begin{equation}
\textrm{stock correlation} \sim \textrm{ownership ties} + \textrm{board ties} + \textrm{research ties}
\end{equation}
As dependent variable we take the pairwise correlation values between all stock time-series and scale them to integer values in the interval $[0,100]$, as the multiplex regression model allows for the dependent variable to be weighted (positive, integer) - See \cite{casiraghi2017multiplex} for details on this regression method. The dependent variables are the unweighted network structures; the Ownership network, the Board network and the Research network. \textit{The regression identifies all dependent variables to be significant in explaining the stock correlations}, see Table \ref{regression_table}.\\

On the basis of the above we conclude that the different layers have distinct information to convey. The three different methods have shown the significance of all layers in the multiplex. These results highlight the different channels of influence between companies have very different structures. The complementary characteristics of the different layers show that the corporate world is an even smaller one than previously reported based on the studies of ownership \cite{battiston2004statistical} or the board membership network \cite{davis2003small}.

\subsection{Network dynamics}

\begin{table}[t!]
\caption{Multiplex network regression on pairwise stock correlations. All layers are found significant independent variables for explaining the stock correlations.}\label{regression_table}
\centering
\begin{tabular}{@{}l cccc @{}}
\toprule
& \multicolumn{2}{c}{German stock correlations}
& \multicolumn{2}{c}{UK stock correlations} \\
\cmidrule(lr){2-3} \cmidrule(lr){4-5}
& Coef. & SE & Coef. & SE \\
\midrule
Board & 0.101 & *** & 0.044 & ***    \\
Ownership &  0.079 & *** &  0.070 & *** \\
Research & 0.112 & *** &  0.195 & *** \\
\midrule[\heavyrulewidth]
\multicolumn{5}{@{}l}{* $p < 0.05$, ** $p < 0.01$, *** $p < 0.001$.}\\
\end{tabular}
\end{table}

We now take the first steps to explorer the dynamics of this multilayer structure, and analyse how the network structure is related to company properties. The small world network that we described before is characterised by the presence of hub nodes. Such core nodes act as connectors of different parts of the network. We calculate the centrality of the nodes within the separate layers, and with the MultiRank centrality \cite{rahmede2017centralities} quantify the node prominence in the multiplex. We regress node centralities on company characteristics like revenue, stock return, revenue growth, and the Sharpe ratio of the stock returns, in an ordinary least squares regression. We find that centrality in the multiplex is significantly correlated to revenue of the companies - see the appendix for more details. Larger companies, as measures by revenue, are more central in the network. The result is not necessarily surprising \cite{vitali2011network}, as larger companies have more resources to form connections, like subsidiaries and R\&D collaborations. 

Now, we look at the position of the companies in the network, related to the stock performances. We measure stock performance with the Sharpe ratio instead of the direct stock returns. This ratio evaluates mean returns compensated for (high) volatility of the stock. A high Sharpe ratio corresponds to a high return and low volatility: a consistent steady high return. The Sharpe ratio is calculated from the returns of a portfolio of stocks $r$ as:
\begin{equation}
    S= \frac{\langle d \rangle}{\sigma_d}, 
\end{equation}
where $d=r_{\textit{portfolio}} - r_{\textit{risk free}}$. We take the risk free return on capital as zero (an assumption which is valid as the return on German governmental bonds is currently zero, or even negative). We calculate the centrality also in the multiplex excluding the stock interactions layer.

Calculating the Sharpe ratio of the portfolio of the highest ranked quantile, using the MultiRank centrality ranking on the multilayer network, we find the core companies perform significantly better compared to the rest of the network. In fact, we find a Sharpe ratio for the portfolio of companies in the core of $0.37$, while only $0.18$ for the rest of the network for the German multiplex ($0.21$  versus $0.07$ for the UK multiplex). However, this effect seems mostly driven by the companies with largest revenue, rather than uniformly by all companies in the core. These preliminary results on the dynamics behind the multiplex structure indicate a relation between company performance and network formation, and invite more research on this topic.  

\section{Conclusion}
In this paper we have described a uniquely compiled dataset which combines various known company-to-company interaction networks into one single multilayer structure. The layers of this system describe different types of interactions between the same set of companies. We have included ownership ties, social ties through joint board members, R\&D collaborations, and stock correlations. 
With three separate methods we show the significance of the multiplex structure. Node and edge overlap highlights that different types of ties connect different sets of players; i.e. the structures are not overlapping and the layers complement each other. 
The structural reducibility quality was used to show that all layers are structurally different and irreducible from an information theory perspective. In the third method we used a regression model to estimate the explanatory power of the multiplex structure on all pairwise stock correlations. The independent variables of network structures of ownership network, board membership network and R\&D network were all found to be significant estimators of the structure in the stock correlations. 

These three methods confirmed that the multiplex representation is different to the single layers or the aggregate, and that these interactions have different structures. For company interactions this indicates studies of peer effects of control should take these multiple connections into account. We evaluate the characteristics of companies related to the multiplex structure. These initial results indicate a relation between company performance and multiplex centrality.
 
Our results show that the corporate world is an even smaller world than the small world already described by various previous studies on corporate control and studies of the 'old boys network' of board rooms. The significance of the different layers of the corporate multiplex invite more research on the interconnectedness of diverse economic and financial networks. 

\appendix
\section{Structural reducibility framework}
The structural reducibility measure evaluates is some layers can be aggregated without loss of distinguishability from the aggregate \cite{de2015structural}. Let our multiplex of four layers be,  $\mathcal{A}=\{A^1,A^2,A^3,A^4\}$, where $A^\alpha$ denotes the adjacency matrix of layer $\alpha$. The Von Neumann entropy of this multiplex is $ H(\mathcal{A}) = \sum\limits_{1}^{4} h_{A^{\alpha}}$, where $h_{A^\alpha}=-\sum\limits_{i=1}^{N}\lambda_i \log_2 (\lambda_i)$, the Von Neumann entropy of $N\times N$ layer $\alpha$ with eigenvalues $\lambda$.
We now sum two or more of the layers find multiplex $\tilde{\mathcal{A}}$. The measure we are maximizing, the relative entropy is $q(\tilde{\mathcal{A}}) = 1-\frac{\mathcal{H}(\tilde{\mathcal{A}})}{h_A}$, where $h_A$ is the entropy of the aggregate of all layers.

\section{Node properties regression}
To establish properties of nodes that can drive the network dynamics we perform a simple regression of company properties to see which of these variables might drive the node ranking in the network. We include company characteristics  revenue, revenue growth, average log return, Sharpe ratio in an ordinary least squares regression:
\begin{equation}
    \textit{node ranking} \sim \langle \textit{revenue} \rangle + \langle \textit{revenue growth}  \rangle + \langle \textit{return} \rangle + \langle \textit{Sharpe ratio} \rangle .
\end{equation}
We find that only revenue is a significant independent value of the node ranking. This results is robust for the multiplex ranking (with and without the stock interactions layer) and with the node centrality (PageRank) from the single layers. 

\section{Applications}
We illustrate the use of this multiplex approach with two further examples of an interactions network of two companies before a partial take-over, and a network structure that shows clustering around audit firm choice. 

\subsection{Interwoven relations before a partial takeover: EON - RWE}
\begin{figure}[ht]
\centering
\includegraphics[width=0.5\linewidth]{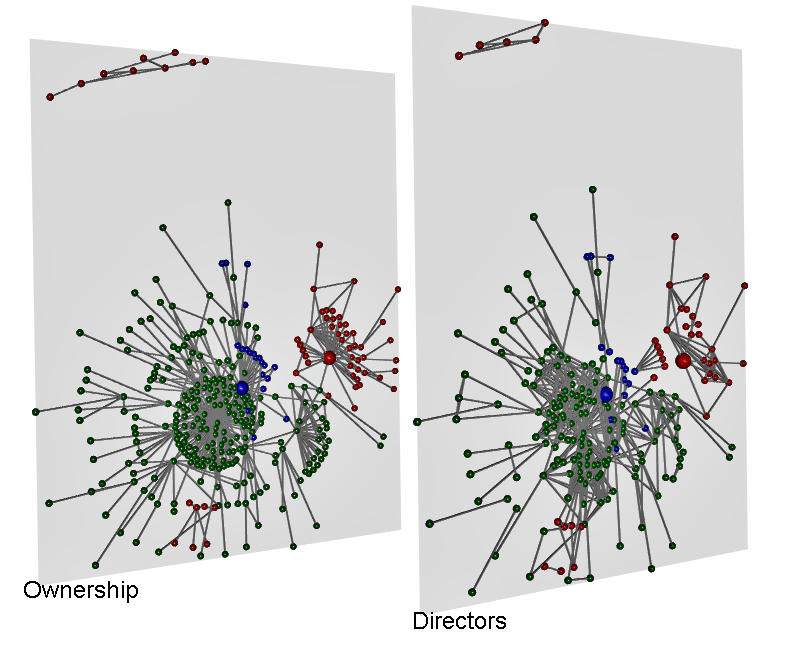}
\caption{The multiplex network of Ownership and Board interactions between German energy conglomerates EON and RWE shows the firms are interconnected well before a partial takeover took place. This network shows relations months before the announcement of a deal in which EON (red) agreed to acquire the Innogy subsidiary (green) of RWE (blue). The two larger nodes are the stock market traded parent entities. Differences in connections in the two visualised layers show certain subsidiaries of EON that are connected to EON by ownership, but that are solely connected through common directors with RWE. Differences in these layers show the insight to gain from multiple types of corporate ties.}
\label{fig:rwe_eon}
\end{figure}

The interactions between two German energy companies reveal different interactions throughout the network. 
Company mergers and takeovers are sometimes preceded by collaborations between the companies. From our network point of view these relations can be visualised, to help understand how interwoven companies are, but also to highlight possible conflicts of interest. As an example of the this we look at the recent deal between German energy conglomerates EON and RWE. In July 2018 the two companies reached a deal where EON will acquire the Innogy subsidiary of RWE, and RWE will end up with a significant stake in EON \cite{eonpressrelease}.
Our multiplex dataset shows a snapshot of the network from early 2018, well before the announcement of this deal. However we can observe a number of directors which are working both for EON and the RWE subsidiary, as well as a number of ownership ties between subsidiaries of the two  companies, see Figure \ref{fig:rwe_eon}. 

\subsection{Clustered auditor choice }
Peer effects, whether through the old boys boardroom network or through herd behaviour, can steer decision making.
All exchange traded companies must have their finances checked every year by an auditor. Big companies usually choose for one of the 'Big Four' audit firms; KPMG, Ernst\&Young, PriceWaterhouseCoopers, and Deloitte. While there is a choice, a recent Financial Times article explains that there are limitations to this free choice \cite{financialtimes_2018}. Audit firms also do consultancy jobs for firms, which presents a conflict for also auditing the books. There are also examples where directors of a company are former partners at auditing firms, and therefore creating a conflict of interest with that auditing firm \cite{financialtimes_2018}. The same goes for the relations between companies that we have described in our multiplex network. 
We can ask whether there is any relation between existing connections between companies and their choice of an auditing firm. With our multiplex network, a simple test is to check whether connected nodes (companies) are more likely to use the same audit firm. For a given audit firm we calculate for all nodes, $n$, the average number of neighbours that use that audit firm. We then compare this average fraction of neighbours which uses an audit firm for all nodes, with the fraction for just the nodes that use that audit firm: 
%\begin{align*}
%\centering
%f_{\text{all nodes}} & = \sum_{\text{all nodes}} \frac{\text{number of neighbors which use audit firm $x$}}{\text{total number of neighbors}}\\
%f_{\text{auditor nodes}} & = \sum_{\text{all nodes which use audit firm x}} \frac{\text{number of neighbors which use audit firm $x$}}{\text{total number of neighbors}}
%\end{align*}

\begin{equation}
    f_{n}  = \sum_{n} \case{\textrm{neighbors with audit firm x}}{\textrm{total number of neighbors}}
\end{equation}
\begin{equation}
    f_{n_{\textrm{auditor}}} = \sum_{n_{\textrm{with audit firm x}}} \case{\textrm{neighbors with audit firm x}}{\textrm{total number of neighbors}}
\end{equation}

We calculate these fractions for each of the big four audit firms individually. 
We find that companies which use the same audit firm are more connected among themselves than to other firms, especially for connections from the Board membership network. This suggests that there is a certain clustering of audit firm choice for connected firms. This effect might be due to other factors, as the co-evolution of the node attribute values (audit firm) and the network structure. The point is that results can be different on different layers of the multiplex and to show how a multiplex network can help in evaluating such questions. 

\begin{figure}[ht]
\centering
\includegraphics[width=0.5\linewidth]{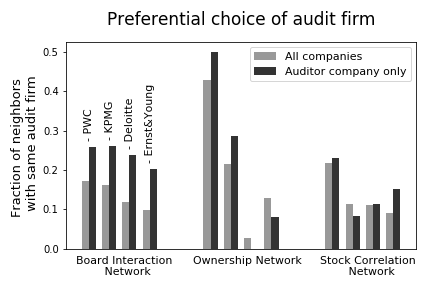}
\caption{Firms that are connected are more likely to use the same audit firm, especially in the Board layer. Shown side by side is the average fraction of neighbours of a node that use the same indicated audit firm; for all nodes and for just the nodes of that that use that audit firm. The higher fraction for nodes of just the auditor indicates that connected firms are more likely to use the same audit firm compared to unconnected firms. There seems to be a relation between o.a. common board members and the choice for a specific audit firm. Such relation is not observed in the the stock interaction network.}
\label{fig:auditor}
\end{figure}

\ack
We thank Giovanni Bonaccorsi for sharing his code for the MultiRank centrality calculations and  
Giona Casiraghi for his code on the multiplex network regression. 

\section*{References}
%\bibliography{sample}

\providecommand{\newblock}{}

\end{document}